\documentclass{aastex62}

\graphicspath{{./}{figures/}}

\received{March 25, 2019}
\submitjournal{ApJ}

\shorttitle{LkCa 15 manuscript}
\shortauthors{Jin et al.}

\begin{document}

\title{New constraints on the dust and gas distribution in the LkCa 15 disk from ALMA}

\correspondingauthor{Sheng Jin}
\email{shengjin@pmo.ac.cn}

\author[0000-0002-9063-5987]{Sheng Jin}
\affil{CAS Key Laboratory of Planetary Sciences, Purple Mountain Observatory, Chinese Academy of Sciences, Nanjing 210008, China}

\author[0000-0001-8061-2207]{Andrea Isella}
\affiliation{Department of Physics and Astronomy, Rice University, 6100 Main St., Houston, TX 77005, USA}

\author[0000-0002-7575-3176]{Pinghui Huang}
\affil{CAS Key Laboratory of Planetary Sciences, Purple Mountain Observatory, Chinese Academy of Sciences, Nanjing 210008, China}
\affiliation{Department of Physics and Astronomy, Rice University, 6100 Main St., Houston, TX 77005, USA}
\affiliation{Theoretical Division, Los Alamos National Laboratory, Los Alamos, NM 87545, USA}
\affiliation{University of Chinese Academy of Sciences, Beijing 100049, Peopleʼs Republic of China}

\author[0000-0002-4142-3080]{Shengtai Li}
\affiliation{Theoretical Division, Los Alamos National Laboratory, Los Alamos, NM 87545, USA}

\author[0000-0003-3556-6568]{Hui Li}
\affiliation{Theoretical Division, Los Alamos National Laboratory, Los Alamos, NM 87545, USA}

\author[0000-0002-9260-1537]{Jianghui Ji}
\affil{CAS Key Laboratory of Planetary Sciences, Purple Mountain Observatory, Chinese Academy of Sciences, Nanjing 210008, China}

\begin{abstract}

We search a large parameter space of the LkCa 15's disk density profile to fit
its observed radial intensity profile of $^{12}$CO (J = 3-2) obtained from ALMA.
The best-fit model within the parameter space has a disk mass of 0.1 $M_{\odot}$
(using an abundance ratio of $^{12}$CO/H$_2$ $=$ 1.4 $\times 10^{-4}$ in mass),
an inner cavity of 45 AU in radius, an outer edge at $\sim$ 600 AU,
and a disk surface density profile follows a power-law of the form $\rho_r \propto r^{-4}$.
For the disk density profiles that can lead to a small reduced $\chi^2$ of goodness-of-fit,
we find that there is a clear linear correlation between the disk mass and
the power-law index $\gamma$ in the equation of disk density profile.
This suggests that the $^{12}$CO disk of LkCa 15 is optically thick and
we can fit its $^{12}$CO radial intensity profile using either 
a lower disk mass with a smaller $\gamma$ or a higher disk mass with a bigger $\gamma$.
By comparing the $^{12}$CO channel maps of the best-fit model with disk models with
higher or lower masses, we find that a disk mass of $\sim$ 0.1 $M_{\odot}$ 
can best reproduce the observed morphology of the $^{12}$CO channel maps.
The dust continuum map at 0.87 mm of the LkCa 15 disk shows 
an inner cavity of the similar size of the best-fit gas model,
but its out edge is at $\sim$ 200 AU, much smaller than the fitted gas disk. 
Such a discrepancy between the outer edges of the gas and dust
disks is consistent with dust drifting and trapping models.

\end{abstract}


\section{Introduction} 

Recently, a large number of protoplanetary disks
were spatially resolved by the
Atacama Large Millimeter/submillimeter Array (ALMA) and the Next Generation Very Large Array (ngVLA)\citep{ALMA2015,Fedele2017,Andrews2018,Isella2018,Liu2018}.
These high-resolution observations of protoplanetary disks 
in dust continuum and molecular line emissions
provide us the morphology of dust and gas distributions in 
a wide variety of protoplanetary disks.
Such information places fundamental constraints for the 
theoretical studies on dust properties and dust-gas interaction,
which is an important building block of planet formation theory
\citep{Dong2015,Isella2016,Jin2016,Liu2018b,Dong2018,Huang2018,Ricci2018,Huang2019}.

LkCa 15 is a 2-5 Myr old K5 star with $L_{\star}$ $\sim$ 0.74 $L_{\odot}$ 
and $M_{\star}$ $\sim$ 1.0 $M_{\odot}$ \citep{Kenyon1995, Simon2000}.
It is located in the Taurus-Auriga star-forming region at a distance 
of 140 pc from the Earth\citep{vandenAncker1998}.
LkCa 15 is an interesting target due to its partially dust-depleted disk 
\citep{Pietu2006, Espaillat2007, Thalmann2010, Andrews2011, Isella2012, Isella2014, Thalmann2014}
and the probability of harboring a planet candidate inside its inner cavity 
\citep{Kraus2012,Sallum2015,Thalmann2016,Currie2019}.
The depleted inner region shown in the dust continuum image 
is about 50 AU in radius
\citep{Pietu2006,Andrews2011,Isella2012,Isella2014},
and it has a mass accretion rate of about $10^{-9}$ $M_{\odot} {\rm yr}^{-1}$\citep{Hartmann1998}.
The outer edge of the dust disk inferred from continuum emission is at $\sim$ 150 AU\citep{Pietu2007,Isella2012}.
The dust mass estimated from 1.3 mm continuum observation is about $5 \times 10^{-4}$ $M_{\odot}$\citep{Isella2012}.
Recent images of scattered light suggest a warped inner disk component inside the inner gap,
providing a clear picture details of the inner gap region of the LkCa 15 disk \citep{Thalmann2015,Thalmann2016,Oh2016}. 

As a young star, LkCa 15 is a luminous source of X-ray and EUV emission\citep{Skinner2013},
indicating that the LkCa 15 disk could still be undergoing the active evolution phase 
of disk physics and chemistry.
Consequently, the protoplanetary disk of LkCa 15 
has been found to be especially 
chemically rich and has been detected in
several molecular transitions\citep{Thi2004, Pietu2007, Chapillon2008, Oberg2010,Punzi2015}.
Molecular line emission shows that the gas disk around LkCa 15 is 
in size of $\sim$ 900 AU\citep{Pietu2007,Isella2012}. 
It is highly optically thick in the emission of $^{12}$CO, 
but optically thin in $^{13}$CO emission\citep{Punzi2015,vanderMarel2015}.
The discrepancy between the outer disk radii shown in the dust continuum and 
molecular line emissions ($\sim$ 150 AU versus 900 AU) suggests that the 
mm-size dust is depleted in the outer part of the LkCa 15 disk,
which is in consistence with dust drifting and trapping models\citep{Birnstiel2010,Pinilla2012}.

Being a protoplanetary disk that may have a young planet candidate,
the LkCa 15 system serves as a unique laboratory 
to study the dust evolution models under planet-disk interaction\citep{Zhu2011,Pinilla2012}.
The distribution of dust and gas in the LkCa 15 disk provides an 
important constraint on theoretic models.
Here, we fit the gas and dust surface density profiles based on
the high-resolution dust continuum image and CO 3-2 line maps obtained from the ALMA. 
We study how the goodness-of-fit changes along with different key
parameters used in the disk density profile, which is related with different physics.
Our disk model and the parameter grid is described in Section \ref{model}. 
In Section \ref{results}, we show the best-fit model with respect to the
observed $^{12}$CO image.
We discuss the dependence of parameters in Section \ref{parameter}.

\section{Observation}

The LkCa 15 was observed at August 17 and August 29, 2014 with ALMA Band 7 (345 GHz, 880 $\mu m$) in ALMA program 2012.1.00870.S (PI P{\'e}rez, L.~M.). 
The $^{12}$CO, $^{13}$CO and C$^{18}$O 3-2 line maps and 0.87 mm dust continuum image were taken with two different tunings.
The $^{12}$CO 3–2 data were obtained with a spectral window centered on 345.796 GHz,
with 1885 channels of 122.06 kHz (0.106 km s$^{-1}$) channel width.
The angular resolution of the $^{12}$CO image is 0.36$\farcs\times$0.23$\farcs$ (50 AU x 32 AU).
The $^{13}$CO 3–2 data were centered on 330.588 GHz, 
with 1967 channels of 122.06 kHz (0.11 km s$^{-1}$) channel width.
The angular resolution of the $^{13}$CO image is 0.28$\farcs\times$0.21$\farcs$ (40 AU x 30 AU).
The C$^{18}$O 3–2 data were centered on 329.331 GHz, 
with 1967 channels of 122.06 kHz (0.111 km s$^{-1}$) channel width.
The angular resolution of the C$^{18}$O image is 0.30$\farcs\times$0.23$\farcs$ (41 AU x 32 AU).
The dust continuum observation consists of four spectral windows(334.01$\sim$335.99 GHz, 339.02$\sim$341.00 GHz, 341.02$\sim$343.00 GHz, 346.01$\sim$347.99 GHz).
Each spectral window has 64 channels of 31248.22kHz (3527.09 km s$^{-1}$) channel width.
The angular resolution of the dust continuum image is 0.23$\farcs\times$0.17$\farcs$ (32 AU x 23 AU).

This is the first comprehensive observation of the CO 3-2 line emission of 
$^{12}$CO, $^{13}$CO and C$^{18}$O in the LkCa 15 disk.
Previous observations revealed the line emission of various molecules
in the LkCa 15 disk,
including the $^{12}$CO 6-5\citep{vanderMarel2015}, CO 2-1 and HCO$^+$\citep{Qi2003,Pietu2007,Punzi2015}, S-bearing molecules\citep{Dutrey2011}, and ethynyl radical (CCH)\citep{Henning2010}.
The $^{12}$CO 2-1 line emission shows that the LkCa 15 disk is highly optically thick in $^{12}$CO\citep{Punzi2015},
and the $^{12}$CO 6-5 emission show that gas is still present inside the observed dust cavity\citep{vanderMarel2015}.
The dust continuum image has also been obtained by several former millimeter observations\citep{Pietu2006,Espaillat2007,Isella2012,Isella2014}, and all these observations show an inner dust cavity of $\sim$ 40-50 AU in size.

\subsection{Dust continuum and line maps}

\begin{figure}
\centering
\includegraphics[width=16.0cm]{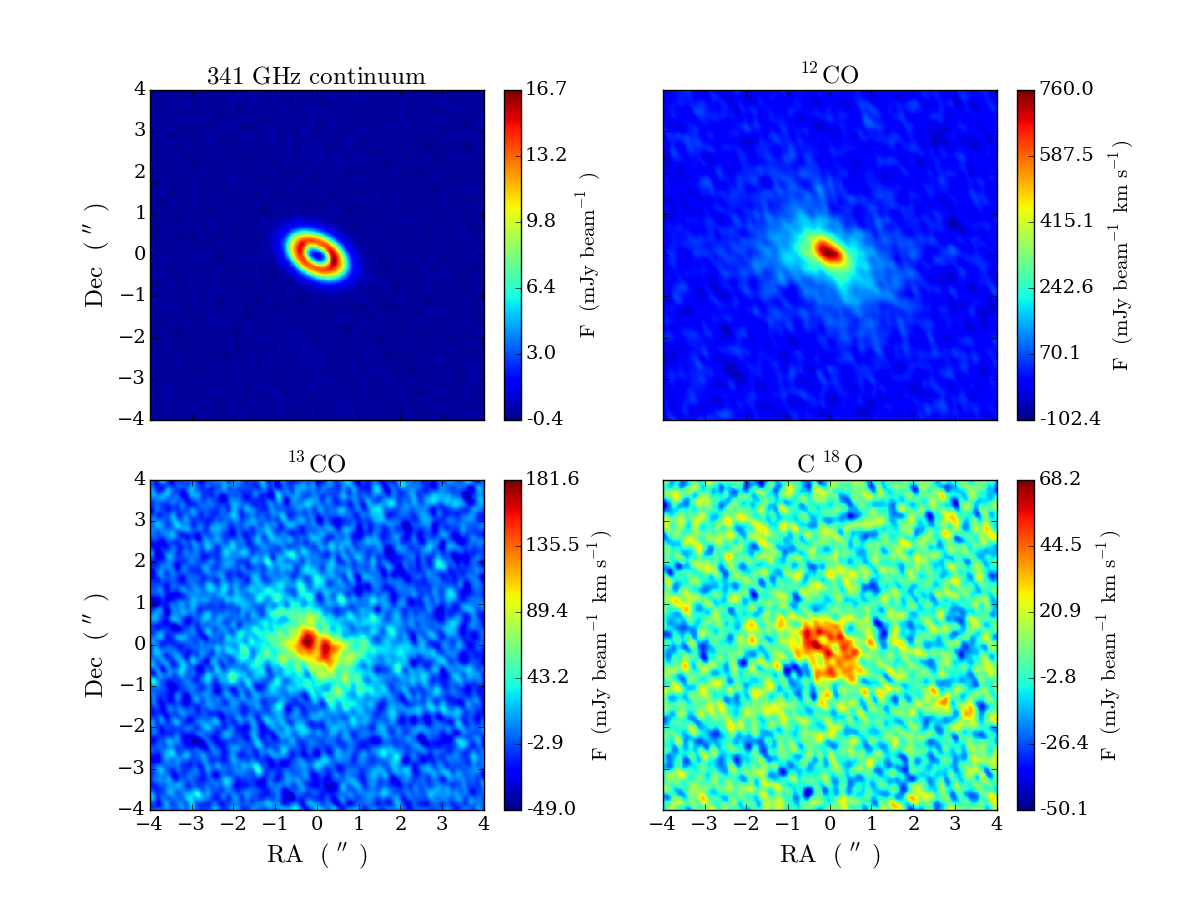}\\
\caption{ALMA observations of the dust continuum, zero-moment $^{12}$CO,
$^{13}$CO and  C$^{18}$O maps of the LkCa 15 system.
Top left: the dust continuum map. Top right: the zero-moment $^{12}$CO map.
Bottom left: the zero-moment $^{13}$CO map. Bottom right: the zero-moment C$^{18}$O map.}
\label{fig1}
\end{figure}

Figure \ref{fig1} shows the dust continuum and 
the $^{12}$CO zero-moment maps of the LkCa 15 system.
We will extract an azimuthal averaged flux from these zero-moment maps along the 
radial direction, and use the derived $^{12}$CO radial intensity profile as a reference
to judge the goodness-of-fit of our disk models.
The dust continuum map shows an inner hole of $\sim$ 65 AU in radius,
and the full width at half maximum of this dust cavity is at $\sim$ 40 AU.
This is in agreement with previous millimeter observations\citep{Andrews2011, Isella2012,Isella2014}.
Such a large inner cavity disappears in the zero-moment map of $^{12}$CO.
For the optically thin $^{13}$CO and C$^{18}$O emissions,
the inner disk show an obvious decrease in the azimuthal averaged flux.
The zero-moment maps of $^{12}$CO, $^{13}$CO and C$^{18}$O
are consistent with previous findings
that the LkCa 15 disk is optically thick in $^{12}$CO emission, 
while optically thin in $^{13}$CO and C$^{18}$O.

Figure \ref{fig2} shows the observed channel maps of 
$^{12}$CO, $^{13}$CO and C$^{18}$O of the LkCa 15 disk.
The channel maps of $^{12}$CO clearly show the near and far halves 
of a double cone structure, which is the feature resulted from a 
circular Keplerian rotational disk\citep{Rosenfeld2013}.
However, we do not observe such a feature in the channel maps of the 
optically thin $^{13}$CO and C$^{18}$O due to lower masses of these two isotopes.

\section{Modeling}
\label{model}

\subsection{surface density profile and parameter grid}

Our primary goal here is to find a gas disk surface density profile that can 
best reproduce the observed zero-moment $^{12}$CO map. 
First, we create a parameter space by parameterizing the 
surface density equation of an analytical disk model.
Then, we search the parameter space to obtain the best-fit model that 
has the least reduced $\chi^2$ of the radial intensity profile of $^{12}$CO emission.

The observed radial intensity profiles of CO isotopes and 
dust continuum emission in the LkCa 15 disk were obtained by
extracting the azimuthal averaged intensity of the zero-moment maps shown in Figure \ref{fig1}.
The radial intensity profiles of CO are extended to
$\sim$ 600 AU, 
while the dust continuum emission shows a ring-shape structure and ends at $\sim$ 200 AU.
The dust continuum and the CO line emissions cannot be fitted
using a single surface density profile.
Thus we use two separate equations to describe the surface density profiles
of dust and gas in the LkCa 15 disk.

We employ parameterized analytical disk surface density profiles to model
the gas and dust in the LkCa 15 disk.
For the gas, we adopt a surface density profile that is described by
\begin{equation}
\label{eq:surfg}
\Sigma_g(r) = \Sigma_0 \Big(\frac{r}{RC}\Big)^{-\gamma} \arctan(r/RC_{\rm arctan})^{\gamma_{\rm arctan}}~~,
\end{equation}
which is a simple power-law density profile combined with
an inner cavity described by an arctangent function.
We turn off the exponential decay term that are typically used to
describe the surface density of protoplanetary disks
\citep[e.g.,][]{Andrews2009}
because the intensity of CO line emissions decrease slowly at larger radii.
Moreover, we fix the $RC$ at 12500 AU to slow down the decrease of the gas density at the outer part of the disk.
There are four free parameters in Equation \ref{eq:surfg}: $\Sigma_0$ that
determines the disk mass, $\gamma$ the power-law decay of the
surface density in the radial direction, $RC_{\rm arctan}$ the size
of the inner cavity, and $\gamma_{\rm arctan}$ the slope of the
junction region between  the inner cavity and the outer disk.
We set up a four-dimensional parameter space of these four free parameters.
The parameter grids at each dimension are listed in Table \ref{para}.

The dust surface density profile is described by
\begin{equation}
\label{eq:surfd}
\Sigma_d(r) = \Sigma_0 \Big(\frac{r}{RC}\Big)^{-\gamma} \exp[-(r/RC)^{2-\gamma}] \arctan(r/RC_{\rm arctan})^{\gamma_{\rm arctan}}
\end{equation}
Compared with the surface density profile of the gas disk, there is an
exponential decay term in the dust density profile
to simulate the disappearance of the dust intensity at larger radii.
Since we have to subtract the dust continuum emission
in generating the $^{12}$CO images for all the models 
in the parameter space listed by Table \ref{para},
we adopt a fixed dust density profile of 
$RC = 66$ AU, $\gamma = -0.15$, $RC_{\rm arctan} = 66$ AU,
$\gamma_{\rm arctan} = 5.35$, and a total dust mass of 
$M_{dust} = 9.8 \times 10^{-5} M_{\odot}$.
These values are determined by fitting the azimuthal averaged radial 
intensity profile of the observed dust continuum map,
and they are used for all the 4096 runs in the parameter space. 

\subsection{physical model}
\label{physicalmodel}

The first step to derive the $^{12}$CO intensity from a specific gas surface density profile
is to calculate the three-dimensional disk temperature structure.
For each gas surface density profile in the parameter space,
We use an iterative approach to obtain a self-consistent
three-dimensional disk temperature structure.
We assume the disk temperature structure is controlled by
micron-size dust particles that are well coupled with gas.
Since the vertical distribution of micron-size dust is
in turn determined by the disk temperature structure,
this is a circular dependency problem.
To solve this problem,
first we generate an initial three-dimensional micron-size dust distribution, and
calculate an initial disk temperature based on this dust distribution.
Using the calculated disk temperature, we produce a new micron-size dust distribution
by solving the differential equation of hydrostatic equilibrium:
\begin{equation}
 \label{hydroequi}
-\frac{\partial \ln \rho_{\rm gas}}{\partial z} =\frac{\partial\ln T_{\rm gas}}{\partial z} + \frac{1}{{c_s}^2}\left[\frac{     G M_\ast z}{(r^2 + z^2)^{3/2}}\right],
\end{equation}
where ${c_s}^2 = k_B T_{\rm gas}/\mu m_h$ is the sound speed.
We then run Monte Carlo radiation transfer simulation to calculate a
new disk temperature structure using the updated dust distribution.
We repeat this process until the three-dimensional disk temperature structure used
in generating a dust distribution and the disk temperature structure calculated by the
same dust distribution converge (the difference between two temperature
structures is within 3\% everywhere).

The disk temperature structure and the dust continuum and CO line maps 
are calculated using the Monte Carlo radiative transfer code 
RADMC-3D\citep{dullemond2012radmc}.
We assume the host star of LkCa 15 is a black-body radiator 
with an effective temperature of 4350 K.
The surface density of micron-size dust in the calculation of the disk temperature profile
is obtained by a dust-to-gas ratio of 0.001.
Compared to the Milky Way average ratio of 0.01 \citep{Bohlin1978},
by setting such a small dust-to-gas ratio we assume that substantial grain growth has occurred in the LkCa 15 disk and the micron-size dust is about 10\% in mass of the total dust.
Submillimeter and dust continuum observations show that dust grain growth to mm-size particles is completed within less than 1 Myr for the majority of circumstellar disks\citep{Rodmann2006, Draine2006, Ricci2010a, Ricci2010b, Ricci2011, Ubach2012}, and simulation of dust evolution also shows mm-size dust can form at the age of $\sim$ 1 Myr at high density region in protoplanetary disk\citep{Ormel2009}.
As aforementioned, the LkCa 15 disk is about 2-5 Myr in age\citep{Kenyon1995, Simon2000}, thus a large fraction of dust can appear to be in large sizes at this stage.
The dust opacities used in this work were calculated similar to that of \cite{Isella2009}. 
We assume the dust grains are compact spheres made of astronomical silicates  \citep{wd01},
organic carbonates \citep{z96}, and water ice, with fractional abundances described in \cite{p94}.
Single grain opacities were averaged on a grain size distribution to obtain the mean opacity.
We adopt a typical MNR power-law size distribution \citep{mnr77}, $n(a) \propto a^{-3.5}$, 
between a minimum grain size of $5 \times 10^{-6}$ mm and a maximum grain size of $1 \times 10^{-2}$ mm.
The resulting dust opacity at the wavelength of 1~$\mu m$ is $5.2 \times 10^3$ cm$^2$ g$^{-1}$.

After we have obtained a self-consistent three-dimensional disk temperature structure,
we calculate the dust continuum and $^{12}$CO line emission.
We interpolate the surface density profile of mm-size dust given by
Equation \ref{eq:surfd} on a three-dimensional spherical grid with
a scale height profile of
$h_{\rm mm-dust}(r) = 0.1 \times 1.0 \textrm{AU} \times (r/20 \textrm{AU})^{1.25}$,
where 0.1 is a parameter that accounts for the settling of 0.15 millimeter size 
dust towards the mid-plane.
This results in a scale height of $\sim$ 0.75 AU at 100 AU for relatively large dust particles. 
Such a scale height is consistent with the findings by \cite{Pinte2015}.
The dust opacity adopted for mm-size dust is calculated using the same model
of the opacity for micron-size dust, the only difference is that 
here the maximum grain size is 1 mm.
The resulting dust opacity at the wavelength of 1~mm is 13.2 cm$^2$ g$^{-1}$.
We convolve the dust continuum map in each model with the PSF of the ALMA observation 
and extract an azimuthal averaged radial intensity profile.
Then we calculate the reduced $\chi^{2}$ for dust continuum emission
between the radial intensity profiles of the model and the ALMA observation.
For different models in our parameter space, the goodness-of-fit of 
the dust continuum emission changes due to the variation of disk temperature. 
But we find that the influence of dust continuum map on
the zero-moment $^{12}$CO map is limited, because the dust continuum image has
a ring-like structure that located at the optically thick region of the $^{12}$CO emission.

To calculate the line emission of $^{12}$CO, we assume the abundance 
ratio of $^{12}$CO/H$_2$ to be 1.4 $\times 10^{-4}$ in mass, and this value is consistent with the canonical ratio of $10^{−4}$ for the disk initial conditions\citep{Lacy1994, France2014}. 
Then we create a three-dimensional density structure of $^{12}$CO based
on the gas surface density profile of each run by solving the 
Equation \ref{hydroequi}.
We include the freeze-out effect of CO by setting the number density of $^{12}$CO
to zero at the region where temperature is below 20 K.
Photodissociation by stellar UV and/or X-ray radiation is another important factor as 
it can destroy the CO in the surface layers of the disk\citep{Visser2009}.
We follow the same procedure as in that of \citet{Qi2011} and \citet{Rosenfeld2013}, 
in which we calculate a photodissociation boundary by vertically integrating the H nuclei density to
a threshold density of 2.0 $\times 10^{20}$. 
This is a mild threshold compared to the value of 5.0 $\times 10^{20}$ used in \citet{Rosenfeld2013}. 
We find that it is hard to fit the $^{12}$CO intensity beyond $\sim$ 400 AU 
using a larger photodissociation threshold like 5.0 $\times 10^{20}$ in H nuclei,
because with such a strong photodissociation rate there are little $^{12}$CO in the outer disk.
The photodissociation threshold turns out to be 
an alternative free parameter in the fitting of the $^{12}$CO radial intensity profile,
as it is a critical parameter for the $^{12}$CO intensity in the outer disk.
In this work, we fix the photodissociation threshold of 
2.0 $\times 10^{20}$ for all the 4096 runs.

We first calculate the self-consistent disk temperature.
Then we calculate the dust continuum map and the channel maps of $^{12}$CO, 
and convolve these images with the corresponding ALMA PSFs.
Afterwards we subtract the dust continuum map from each channel map,
and integrate all the channel maps to generate a zero-moment map of $^{12}$CO.
Finally, we subtract an azimuthal averaged radial intensity profile from 
the zero-moment $^{12}$CO map and calculate the corresponding reduced $\chi^2$ of $^{12}$CO emission.

In addition, once we obtain the best-fit parameter sets for the $^{12}$CO emission,
we tentatively adjust the abundance ratio of $^{12}$CO/$^{13}$CO to fit the radial intensity profile of 
$^{13}$CO, using the same disk temperature profile given by the best-fit $^{12}$CO model.
The observed $^{13}$CO radial intensity profile shows a peak at $\sim$ 45 AU, 
although there is no such a peak in the radial intensity profile of $^{12}$CO (the radial intensity profiles of $^{12}$CO
and $^{13}$CO can be found in Section \ref{results}). 
Hence, we have to adopt different abundance ratios of $^{12}$CO/$^{13}$CO for the disk region inside or outside of 45 AU.
Although the fitting process of $^{13}$CO is not a self-consistent approach, 
the result can at least be an estimation of the number density of $^{13}$CO.
We do not model the C$^{18}$O disk because the signal-to-noise ratio is low and no flux are detected outside of 200 AU.

\section{the best-fit model}

\label{results}

\subsection{$^{12}$CO intensity profile}

The best-fit model in our parameter space has the following parameters:
$M_{\rm disk} = 0.1 M_{\odot}$, $\gamma = 4.0$, $RC_{\rm arctan} =  45$ AU,
and $\gamma_{\rm arctan} = 10$.
Although the fitted disk mass is relatively large, it agrees
with previous findings\citep{Isella2012,Huang2017}.
Figure \ref{fig3} shows the gas surface density profile of this best-fit model,
which is related to a power-law disk without an exponential decay term
usually used in the description of protoplanetary disks \citep[e.g.,][]{Andrews2009}.
Figure \ref{fig4} gives the self-consistent temperature structure as calculated 
by our iterative approach that was described in Section \ref{physicalmodel}.
It shows a typical two-layer vertical structure of passive 
irradiated circumstellar disks \citep{Dullemond2002}.
In the surface layer, the temperature decreases from $\sim$ 51 K at 100 AU to 
$\sim$ 29 K at 500 AU. In the mid-plane, the temperature decreases 
from $\sim$ 15 K at 100 AU to $\sim$ 11 K at 500 AU.

The top panel of Figure \ref{fig5} compares the azimuthal averaged radial intensity profiles
extracted from the zero-moment $^{12}$CO maps of our best-fit model with the ALMA observation.
The best-fit model results in a reduced $\chi^2$ of 2.51.
Its  gas surface density profile has an inner cavity of 45 AU in size.
However, we do not see such a cavity in the resulting radial intensity profile,
since most of the inner cavity is optically thick in $^{12}$CO emission.
In fact, for a majority part of 
the gas density profiles in our parameter space, the inner disk region is optically thick in 
$^{12}$CO emission and its intensity only depends on the calculated disk temperature.
Although the very inner part of the disk cavity can become optically thin 
due to the arctangent function used in the surface density equation,
this inner optically thin region did not show up in the 
radial intensity profile because of the large PSF of the ALMA observation, 
which is of $\sim$ 50 AU in major axis and $\sim$ 32 AU in minor axis.
The observed $^{12}$CO intensity decreases slowly in the 
outer optically thin part of the disk.
We find that it is hard to fit the slope of the $^{12}$CO radial intensity profile 
in the entire disk using a simple power-law surface density.
Our best-fit model has a large $\gamma$ of 4.0.
It fits the inner $\sim$ 400 AU very well, but beyond $\sim$ 400 AU
it shows lower intensity compared to the ALMA observation.
Note that the photodissociation threshold can be another free parameter in 
our model that can affect the intensity in the outer part of the disk.
In order to slow down the decrease of the $^{12}$CO intensity beyond $\sim$ 400 AU,
we set a photodissociation threshold of 2.0 $\times 10^{20}$ in H nuclei
to keep more $^{12}$CO in the outer disk,
which is a mild threshold compared to the value of 5.0 $\times 10^{20}$ used in \citet{Rosenfeld2013}.
Different photodissociation thresholds will result in different best-fit parameter sets.

\begin{figure*}
\centering
\includegraphics[width=15.0cm]{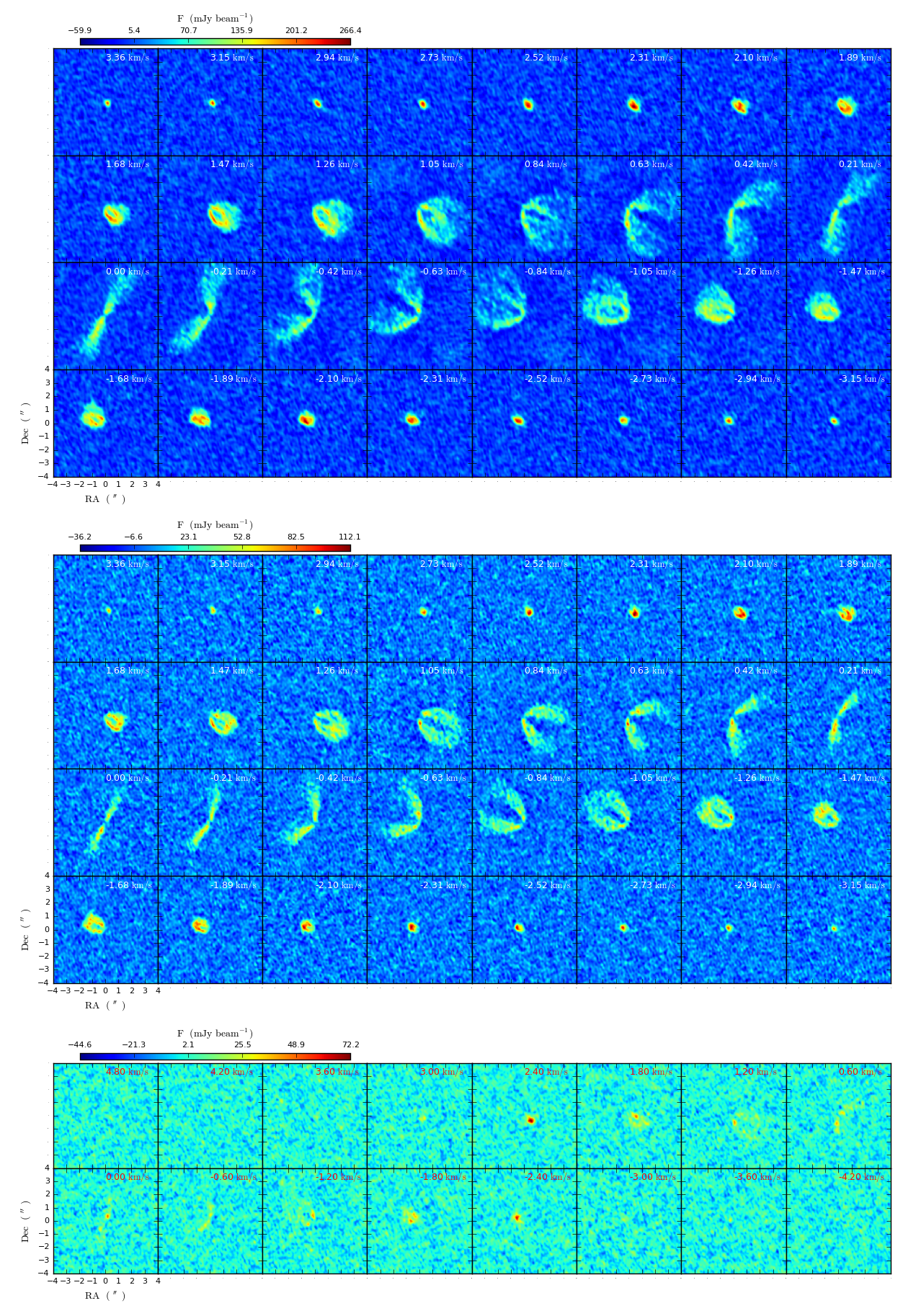}
\caption{
The observed channel maps of $^{12}$CO, $^{13}$CO and  C$^{18}$O of the LkCa 15 system.}
\label{fig2}
\end{figure*}

\begin{figure*}
	\centering
\includegraphics[width=14.0cm]{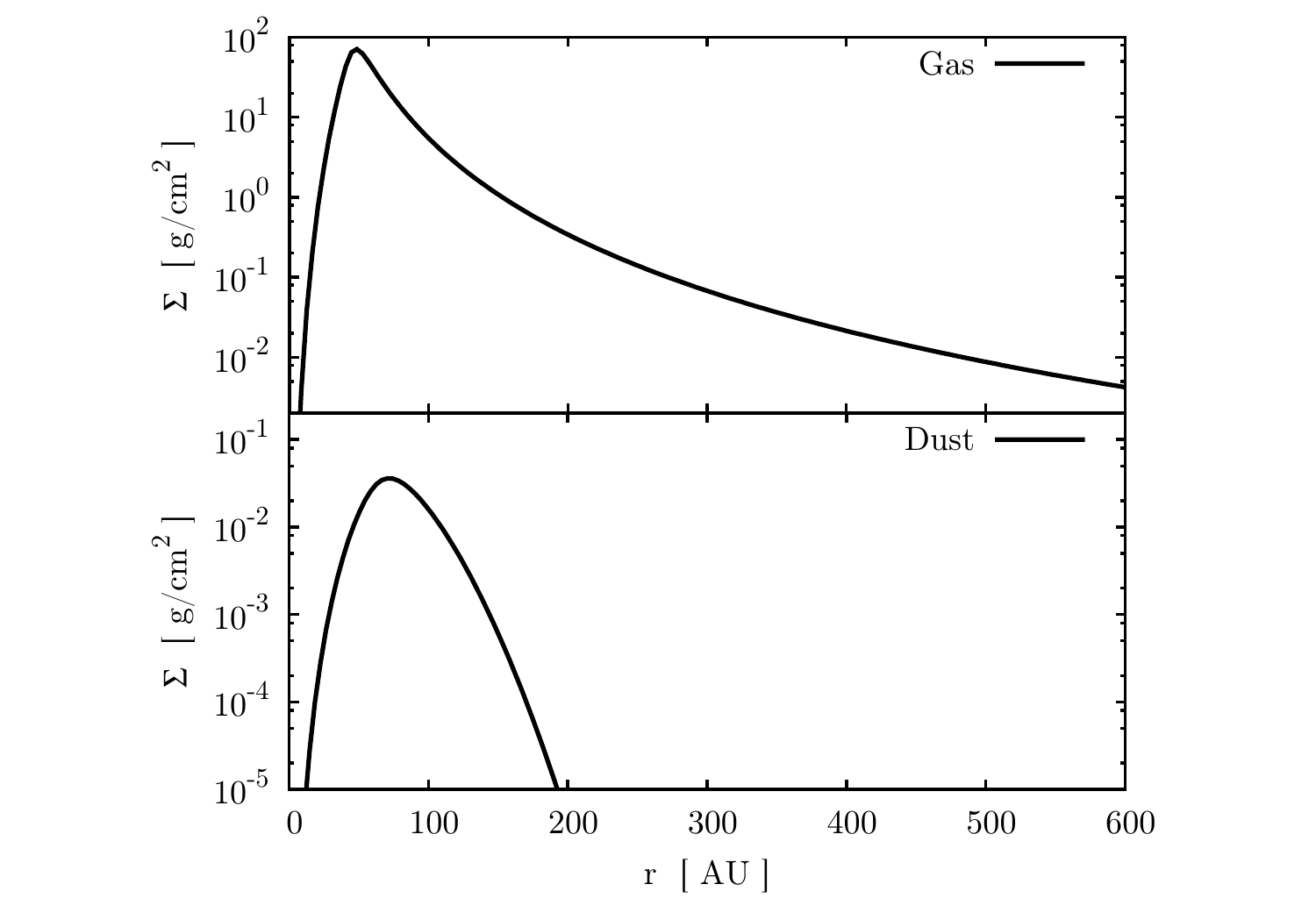}\\
\caption{
Gas surface density profile of the best-fit run (top panel) and
the surface density profile of mm-size dust 
used in all the 4096 runs (bottom panel).
}
\label{fig3}
\end{figure*}

\begin{figure*}
	\centering
\includegraphics[width=14.0cm]{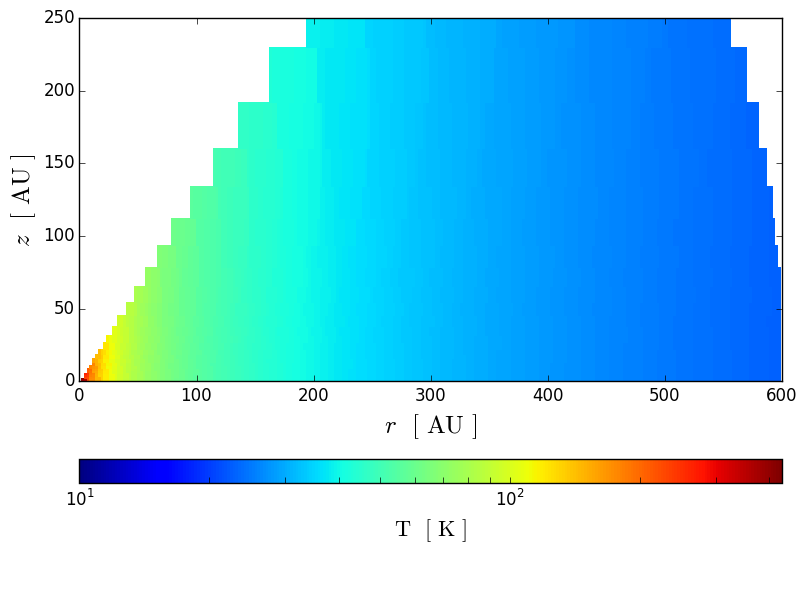}\\
\caption{
The vertical disk temperature structure of the best-fit model
as calculated by our iterative approach.
}
\label{fig4}
\end{figure*}

\begin{figure*}
	\centering
\includegraphics[width=14.0cm]{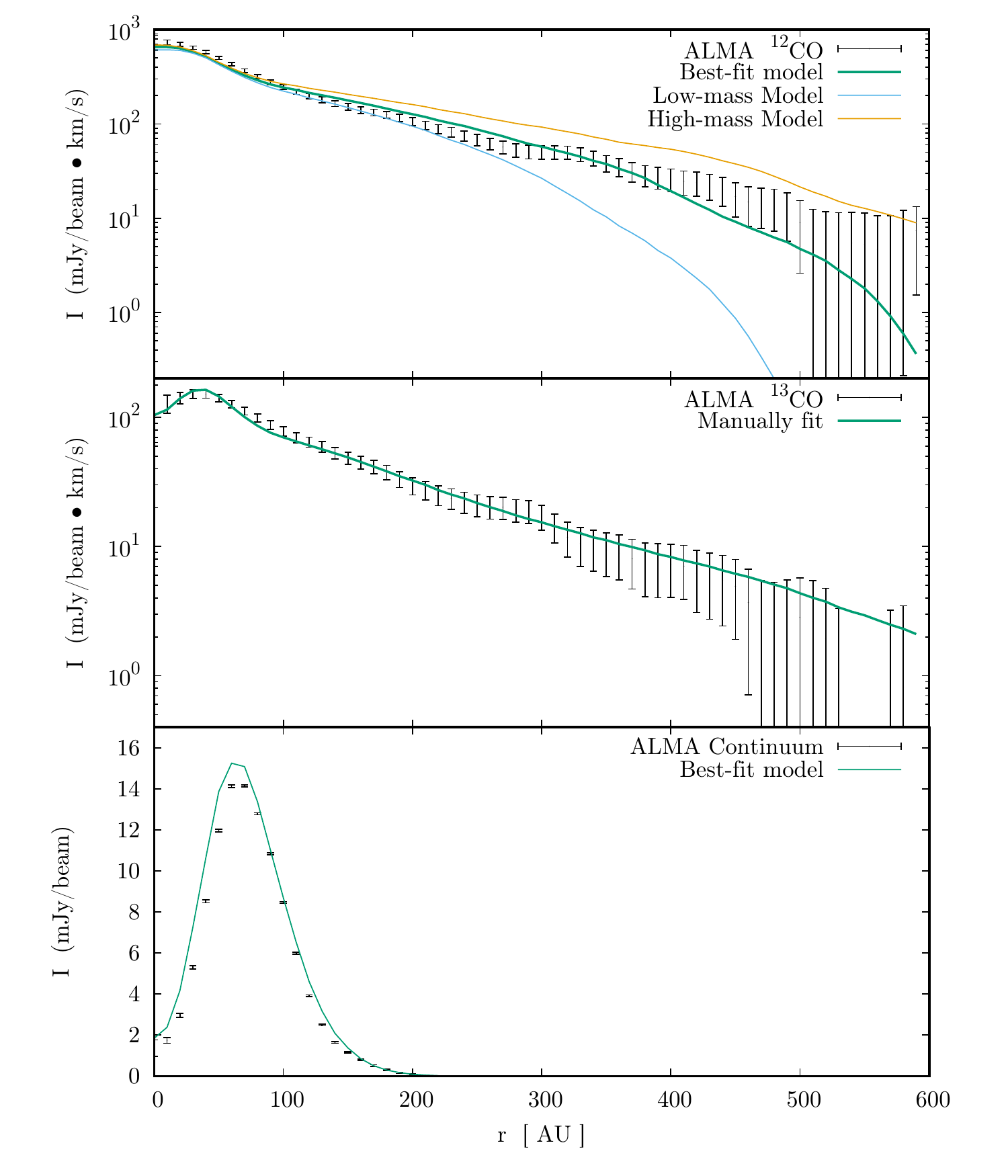}\\
\caption{
Comparison of the azimuthal averaged radial intensity profiles that are
extracted from the zero-moment maps of $^{12}$CO, $^{13}$CO 
and the dust continuum map.
The error-bars show the ALMA observation,
and the solid green lines show our our best-fit model.
The brown and blue lines show the intensity profiles of the high-mass and low-mass model
	in the parameter study at \ref{channelmapmorph}. 
}
\label{fig5}
\end{figure*}

\begin{figure*}
	\centering
\includegraphics[width=16.0cm]{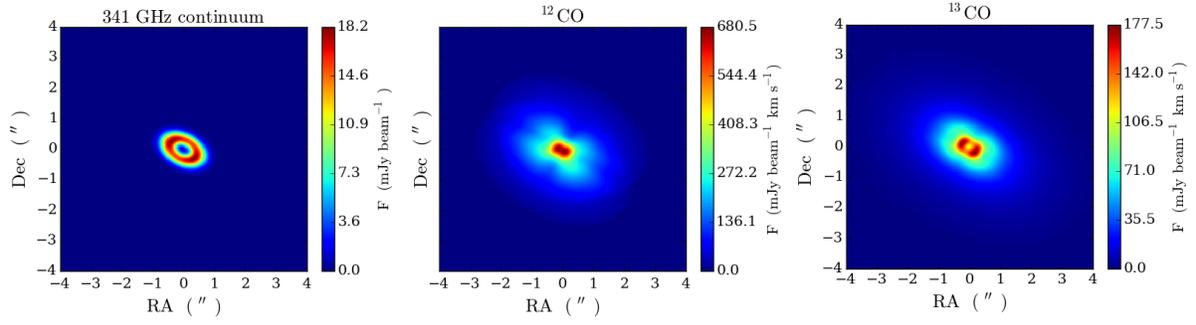}\\
\caption{The dust continuum and zero-moment $^{12}$CO of our best-fit model and 
the zero-moment map of the fitted $^{13}$CO.
Left: the dust continuum map. Middle: zero-moment $^{12}$CO map.
	Right: zero-moment $^{13}$CO map.}
\label{fig6}
\end{figure*}

\begin{figure*}
	\centering
\includegraphics[width=12.0cm]{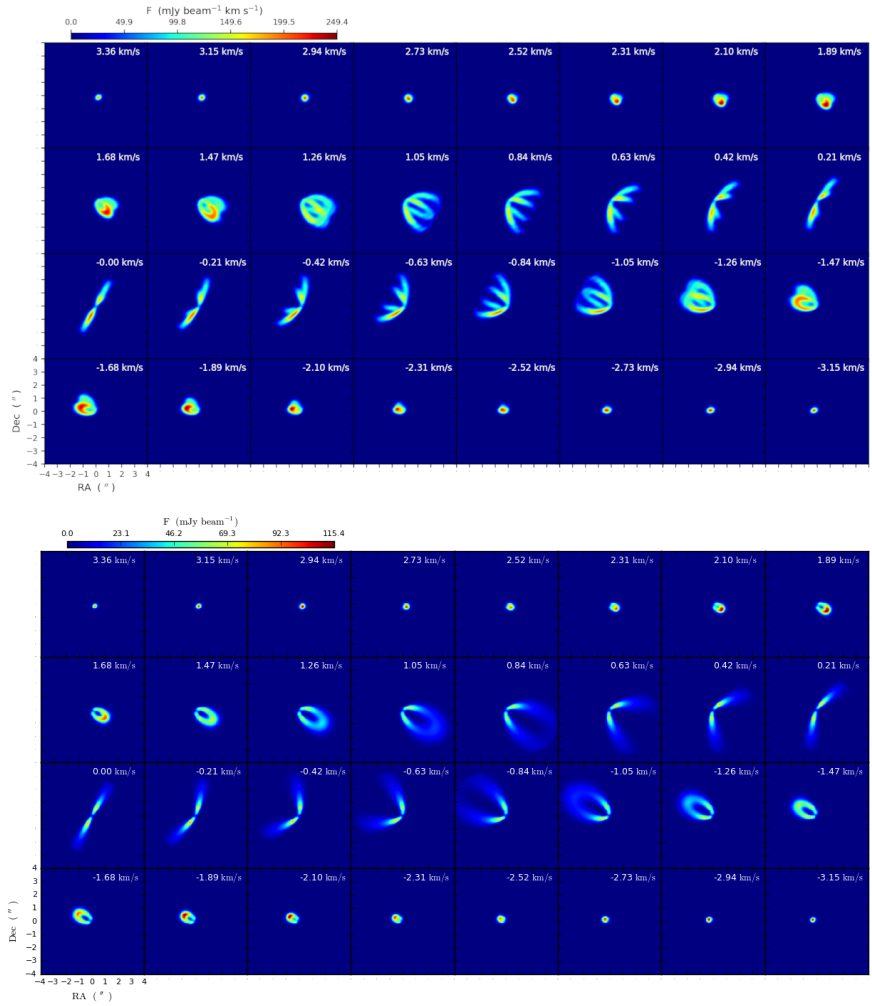}
\caption{
The channel maps of $^{12}$CO of the best-fit model and the channel maps of the fitted $^{13}$CO.}
\label{fig7}
\end{figure*}

In Figure \ref{fig6} and \ref{fig7} we show the dust continuum 
map, the $^{12}$CO zero-moment map, and the $^{12}$CO channel maps
of our best-fit model.
Rather than interpolating the Fourier transformation of our model images 
to the actual observation dataset and cleaning the dataset to get exactly the same PSF of ALMA,
we simply convolve a Gaussian function of the same sizes of major 
and minor axes with the ALMA observation to speed up our fitting process. 
Thus, compared to Figure \ref{fig1} and \ref{fig2},
our model images are more smooth than the ALMA observation.
For the calculation of reduced $\chi^2$,
we only use the azimuthal averaged intensity profile.
We notice that the channel maps provide important constraints on the goodness of fitting.
In the $^{12}$CO channel maps shown in Figure \ref{fig7}, we can 
clearly see the near and far halves of a double cone structure of a Keplerian disk,
and the relative angle and magnitude of the near and far halves share 
similarities with the observed structure shown in Figure \ref{fig2}.
These similarities between the channel maps of our best-fit model 
and the observed channel maps suggest that 
the mass of LkCa 15 disk is around 0.1 $M_{\odot}$ under the assumption
that the abundance ratio of $^{12}$CO/H$_2$ is $\sim$ 1.4 $\times 10^{-4}$ in mass.
The fitting of the outer disk is not good as the disk size is smaller in the 
channel maps shown in Figure \ref{fig7}, while the 
observed channel maps in Figure \ref{fig2} exhibits more extended images. 
This difficulty is due to the simple power-law surface density profile and the 
identical radial photodissociation threshold used in our model.
We will investigate the effect of different disk masses in Section \ref{channelmapmorph}. 
Note that the fitted disk mass depends on the abundance ratio of $^{12}$CO/H$_2$ 
in our model.
If we adopt a larger abundance ratio of $^{12}$CO/H$_2$ of
$\sim$ 3.0 $\times 10^{-4}$ in mass, 
the fitted mass of the LkCa 15 disk should be around 0.05 $M_{\odot}$.

The bottom panel of Figure \ref{fig5} compares the azimuthal averaged radial intensity profile
of the dust continuum emission of our best-fit model with the ALMA observation.
The observation bias for dust continuum image is small, as shown by
the small error-bar in the observed radial intensity profile.
As a result, the radial intensity profile extracted from the
dust continuum image of our best-fit model has a large reduced $\chi^2$ of 255.
But the two intensity profiles of the best-fit model
and the ALMA observation generally match.
The $\chi^2$ of dust continuum emission can be largely reduced by 
further fine tuning the dust surface density profile.

\subsection{$^{13}$CO image}

Based on the surface density profile and the temperature structure of 
the best-fit model of $^{12}$CO, we manually adjust the 
abundance ratio of $^{12}$CO/$^{13}$CO to fit the observed $^{13}$CO intensity.
This actually exhibits the $^{13}$CO emission at the temperature profile given by the gas density profile obtained by the fitting of $^{12}$CO,
and it is not a self-consistent way as compared to the fitting of $^{12}$CO intensity.
We adopt this simplified approach only to estimate the mass of $^{13}$CO needed to 
reproduce the observed $^{13}$CO image based on the disk properties of the fitted $^{12}$CO disk.

We use a power-law function to describe the radius-dependent 
abundance ratios of $^{12}$CO/$^{13}$CO:
\begin{equation}
\label{eq:isotp}
\eta(r) = n_0 \Big(\frac{r}{45AU}\Big)^{\alpha} ~~, 
\end{equation}
where $\eta(r)$ is the abundance ratio at $r$, and
$n_0$ is the abundance ratio at 45 AU.
We separately fit the $\alpha$ for the inner and outer disk regions
by manually adjusting $n_0$ and $\alpha$.
We obtained a $n_0$ of 6360,
$\alpha$ = -4.0 for the disk region inside of 45 AU,
and $\alpha$ = 1.7 outside of 45 AU. 
This leads to $^{12}$CO/$^{13}$CO = 1, 248, and 6360 at 5, 20, and 45 AU,
and $^{12}$CO/$^{13}$CO = 1640, 500, 155 and 78 at 100, 200, 400 and 600 AU.
Since the $\gamma$ in the best-fit $^{12}$CO disk is 4, 
we can infer that the number density profile of
$^{13}$CO is of the form $r^{-2.3}$ outside of 45 AU.
This could partly explain why it is difficult to fit the $^{12}$CO intensity in the outer disk
using a surface density profile that is of the form $r^{-4}$,
given that the number density profile fitted from the optically thin $^{13}$CO 
is of $r^{-2.3}$.
But if we use a surface density profile that is of the form  $r^{-2.3}$,
the fitted disk mass will be around $10^{-3}$ to $10^{-2}$ $M_{\odot}$ (see Section \ref{parastudy} for details).
Such a low mass can not reproduce the morphology of the observed $^{12}$CO channel maps
as shown in Section \ref{channelmapmorph}.
Thus it is difficult to fit the $^{12}$CO intensity in the inner and outer disk regions
using a simple power-law radial density profile.
Figure \ref{fig5} shows the radial intensity profile of $^{13}$CO 
compared with the ALMA observation.
The reduced $\chi^2$ of $^{13}$CO is derived to be 0.33.

The zero-moment and channel maps of $^{13}$CO obtained from 
the aforementioned fitted abundance ratios are shown in Figure \ref{fig6} and \ref{fig7}.
The fitted  zero-moment map of $^{13}$CO has an inner cavity,
which means the disk becomes optically thin at the wavelengths of $^{13}$CO line emissions.
The observed zero-moment map of $^{13}$CO does not show a distinct inner cavity, 
but we can see an obvious decrease of the intensity in the inner disk region.
For the channel maps, since the mass of $^{13}$CO is much lower than the $^{12}$CO,
we do not see the near and far halves of a double cone structure
as shown in the channel maps of $^{12}$CO.
The morphology of the $^{13}$CO channel maps agrees well with the observation.

\section{a parameter study}
\label{parameter}

\subsection{dependence of parameters}
\label{parastudy}

\begin{figure*}[!htb]
	\centering
\includegraphics[width=17.0cm]{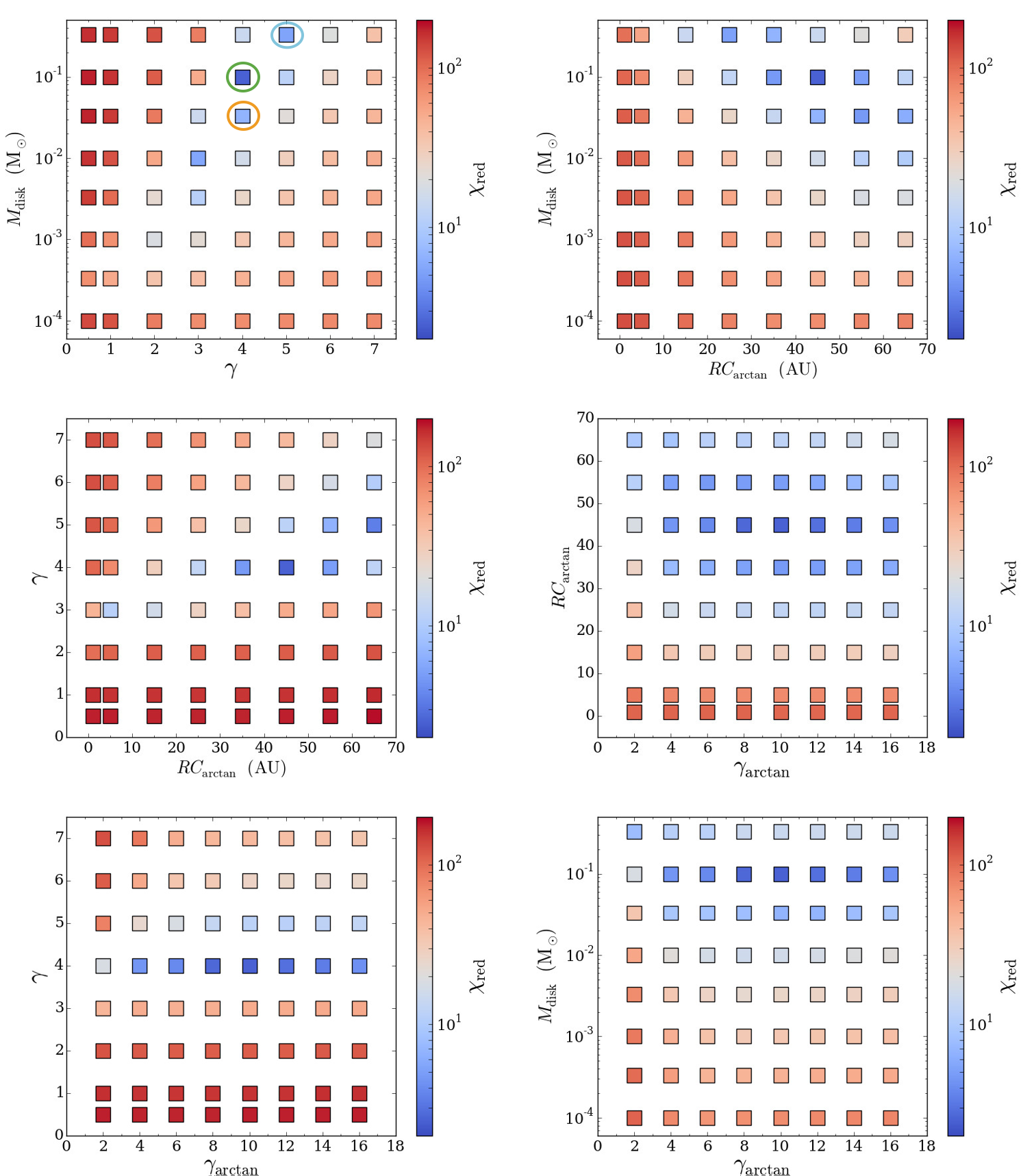}
\caption{
Heat maps of the reduced $\chi^2$ of $^{12}$CO in different combinations of parameters.
}
\label{fig8}
\end{figure*}

We aim to investigate how the fitting of the $^{12}$CO intensity changes with different parameters.
For this, we plot in Figure \ref{fig8} the heat maps of the reduced
$\chi^2$ of the models that have at least two equal parameters
with our best-fit model.
For example, The top left panel shows the 64 models
that have the same $RC_{\rm arctan}$ and $\gamma_{\rm arctan}$
as the best-fit model (marked with a green circle).
We see that for the models that have relatively lower reduced $\chi^2$,
where they show a clear linear correlation between the $M_{\rm disk}$ and $\gamma$.
This means that we can find a reasonable fit of the observed $^{12}$CO
intensity using either a low disk mass with a small $\gamma$ or 
a high disk mass with a large $\gamma$.
This relation can be explained as follows.
For most of our 4096 models, the inner part of the disk (inside of $\sim$ 300 AU) is optically thick for the line emission of $^{12}$CO.
As a result, the disk temperature profile determines the observed intensity
in the inner part of the disk.
Since all 64 runs in this panel have the same $RC_{\rm arctan}$ and 
$\gamma_{\rm arctan}$, they will have similar disk temperature structures in the inner part of the disk.
Thus the key factor to obtain a good fit 
for these 64 runs is to derive the right intensity in the outer disk region.
This means that in the case of a high mass disk,
we only need to reduce the mass at the outer disk region to obtain low intensity there
to fit the observed $^{12}$CO emission.
According to Equation \ref{eq:surfg}, to achieve this point, 
we can set a large $\gamma$ to decrease the disk surface density quickly at large radii.
In Figure \ref{fig5} we also plot the intensity profiles of two disks 
that has a higher disk mass of of 0.33 $M_{\odot}$ or 
a lower disk mass of 0.033 $M_{\odot}$,
as compared to the best-fit model.
The intensity profiles of these three models in the inner 100 AU overlap each other.
The high-mass model result in a reduced $\chi^2$ of 14.6,
and the low-mass model result in a reduced $\chi^2$ of 6.7.
The difference between three models is the goodness-of-fit at the outer disk region.
We will show in Section \ref{channelmapmorph} that the goodness of these three models
can also be clearly observed in the morphology of the $^{12}$CO channel maps.

The top right panel of Figure \ref{fig8} shows the 64 models
that have the same $\gamma$ and $\gamma_{\rm arctan}$ with our best-fit model.
This panel shows that for the disks that have a small inner cavity 
(with size $<$ 10 AU), they cannot reproduce the observed $^{12}$CO intensity.
Only models that have an inner cavity of $\sim$ 30-60 AU can possibly
obtain a small reduced $\chi^2$.
Furthermore, we can see in the case that the $\gamma$, $\gamma_{\rm arctan}$ 
and $M_{\rm disk}$ are fixed, the goodness-of-fit shows weak correlation 
to the size of the inner cavity ($RC_{\rm arctan}$).
On the other hand, when the $\gamma$, $\gamma_{\rm arctan}$ and $RC_{\rm arctan}$ are fixed, 
the goodness of fitting shows weak correlation to the disk mass, since the $^{12}$CO emission is optically thick.

The middle left panel of Figure \ref{fig8} shows the 64 models
that have the same $M_{\rm disk}$ and $\gamma_{\rm arctan}$ with the best-fit model.
The row with $\gamma$ = 4 shows again that the size of the inner cavity 
plays a less important role in the goodness of fitting. 
The essential part in the fitting of the $^{12}$CO intensity is to find a combination of 
$\gamma$ and $M_{\rm disk}$ that can fit the outer part of the disk.
If this goal is achieved, then we can obtain a model that fit the observation,
regardless small changes in $\gamma_{\rm arctan}$ and $RC_{\rm arctan}$.

The middle right panel of Figure \ref{fig8} shows how the reduced $\chi^2$ depends
on $\gamma_{\rm arctan}$ and $RC_{\rm arctan}$, i.e., the size of the inner cavity 
and the slope of the connection region between the inner cavity and outer disk.
It confirms that these two parameters have weak effect 
on the goodness-of-fit of the $^{12}$CO intensity.
However, there should be a large inner cavity.
As it shows that an inner cavity of size $<$ 20 AU do
not reproduce the observed $^{12}$CO intensity.

The bottom two panels of Figure \ref{fig8} shows how the reduced $\chi^2$ changes
in the $\gamma$ versus $\gamma_{\rm arctan}$ space and the $M_{\rm disk}$ versus $\gamma_{\rm arctan}$ space.
The left panel shows that if the $M_{\rm disk}$ is fixed, then the $\gamma$ determines the goodness-of-fit. 
On the contrary, the right panel shows that if the $\gamma$ is fixed, the $M_{\rm disk}$ determines the goodness-of-fit.
For the $\gamma_{\rm arctan}$, here we see again that it has a very limited effect on the goodness-of-fit.
Therefore, we may conclude that the most important part in obtaining a good fit is to find a combination of $M_{\rm disk}$ and $\gamma$.

\subsection{Constraint on the disk mass}
\label{channelmapmorph} 

The best-fit model in our parameter space has a disk mass
of 0.1 $M_{\odot}$, a $\gamma$ of 4.0, and an inner cavity of 45 AU.
We have seen in Section \ref{parastudy} that the goodness-of-fit is affected 
by different combination of parameters, 
as the top left panel in Figure \ref{fig8} shows that by adjusting the $\gamma$, 
we can obtain reasonable fits with  higher or lower disk masses.
Here we investigate how much our model constrains the mass of the LkCa 15 disk.
We choose three models from the top left panel in Figure \ref{fig8}, the best-fit model, a high-mass disk model of 0.33 $M_{\odot}$
and a low-mass model of 0.033 $M_{\odot}$ (these three models are marked with a green, brown and blue circle respectively).
The radial intensity profiles of three models shown in Figure \ref{fig5}, and the high-mass
or low-mass model either over-produce or under-produce the intensity out of $\sim$ 100 AU.
Since the radial intensity profile is extracted from the zero-moment map
that is actually a degenerated image obtained by combining all the channel maps in an actual observation, 
we expect the difference of these models can also be observed in the channel maps.

\begin{figure*}[!htb]
	\centering
\includegraphics[width=15.0cm]{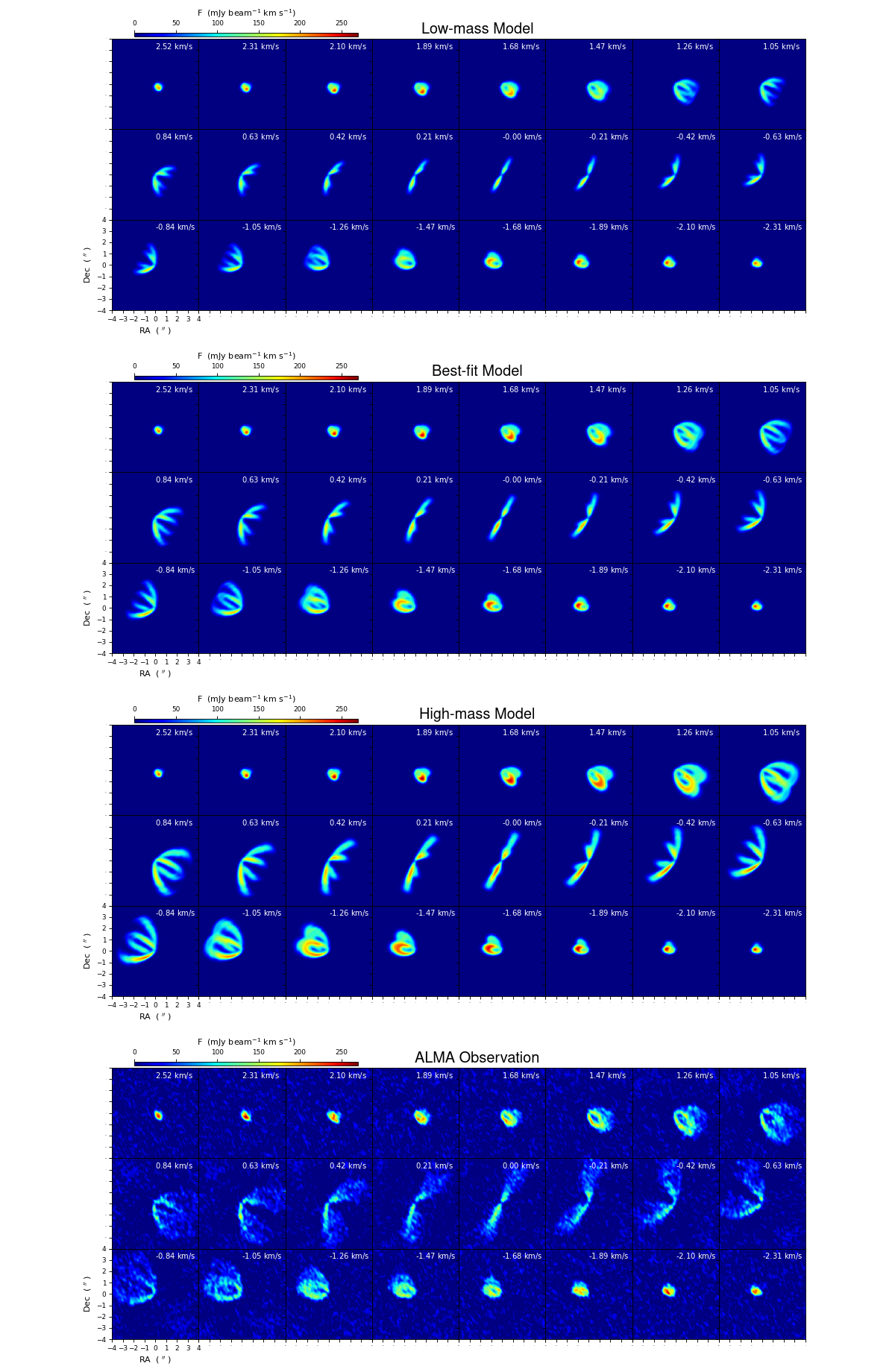}
\caption{
The $^{12}$CO channel maps of the low-mass, best-fit, high-mass model and the ALMA observation.
}
\label{fig9}
\end{figure*}

In Figure \ref{fig9} we compare the three models' channel maps with the ALMA observation, where all the channel maps are plotted in the same colorbar.
The most apparent difference between three models is the size of the channel maps. 
Since we have seen in Figure \ref{fig5} that the high-mass and low-mass model result in 
either higher or lower intensity in the outer part of the disk,
they either show larger or smaller size in the channel maps compared with the best-fit model.
There is another feature that can be used as an effective criteria for the goodness-of-fit of the disk mass,
i.e., the morphology of the near and far halves of a double cone structure of a Keplerian disk.
For example, in the channel map at the velocity of $\pm$ $1.05 km/s$, 
the high-mass model shows a too much intense far halve in the double cone structure 
compared to the observation,
while the low-mass model shows a much smaller double cone structure since it under-produce the intensity at the out disk region.
In the channel maps at the velocity of $\pm$ $0.63$ and $\pm$ $0.42 km/s$, 
the two short wings of the double cone structure in the high-mass 
model are too strong compared to the observation.
Thus, the inability of the high-mass and low-mass model in fitting 
the intensity in the outer disk region
also exhibits in the morphology of the double cone structure in the channel maps.
They show a double cone structure that is either too strong or too weak compared to the 
observed channel maps.

The mm-size dust surface density profile that can reproduce the observed dust continuum map is
significantly different compared to the fitted gas surface density profile, 
and this is a reliable result because of the 
small observational bias of the dust continuum emission.
The mm-size dust density profile has a peak at $\sim$ 65 AU, 
indicating a pressure bump exists at the same location in the gas disk.
The full width at half maximum of this dust cavity is at $\sim$ 40 AU,
similar as the inner cavity of the fitted gas disk.
The mm-size dust disk has an outer edge at $\sim$ 200 AU, 
which is much smaller compared to the fitted gas disk of $\sim$ 600 AU in size.
Such a discrepancy between the gas and dust disks provides an important constraint of
the dust drifting models in the LkCa 15 system.
Our mm-size dust disk is of $\sim$ 1.0 $\times 10^{-4}$ $M_{\odot}$ in mass.
According to our dust opacity model, the opacity at mm wavelength
is mainly contributed by dust of $\sim$ 0.15 mm is size.
This suggests a dust-to-gas ratio of $\sim$ 0.001 for dust of $\sim$ 0.15 mm in size.
But this is a weak constraint because we adopt a uniform 
opacity model for dusts at different distances.
In reality, the species and the size distribution of dust should change
at different radii, and this will affect the calculated dust opacity and 
consequently the fitted dust masses at different radii.

\section{Conclusions}
\label{conclu}

In this work, we analyze the dust continuum and $^{12}$CO 3-2 line emission
maps of the LkCa 15 disk that are obtained from ALMA observation.
We parameterize an analytical surface density profile of the LkCa 15 disk
and search through the parameter space to find the best-fit 
gas surface density model that 
leads to the least reduced $\chi^2$ of the $^{12}$CO intensity.
Our key findings are summarized as follows:

1. The best-fit model of the gas disk based on $^{12}$CO 3-2 line emission is a disk
of 0.1 $M_{\odot}$ in mass. The gas disk has an inner cavity of 45 AU in size, 
and its outer edge is at $\sim$ 600 AU.
The surface density profile of this best-fit model follows 
a power-law of the form $\rho_r \propto r^{-4}$.
But such a steep power-law density profile results in lower
$^{12}$CO intensity at outer part of the disk beyond $\sim$ 400 AU
compared to the observation.

2, The dust continuum map can be reproduced by a dust disk which has 
an inner cavity of $\sim$ 65 AU in size, and 
the full width at half maximum of this cavity is $\sim$ 40 AU.
The size of the dust cavity is similar to the size of the fitted gas cavity.
Unlike the gas disk, the mm-size dust disk ends at $\sim$ 200 AU.
The discrepancy between the outer edges of the gas and dust
disks can be used to study the dust drifting models in the LkCa 15 disk.

3, The heat maps of the reduced $\chi^2$ of different models 
show a linear correlation between the 
$M_{\rm disk}$ and $\gamma$ for the models that have a reasonable goodness-of-fit
of the radial intensity profile of $^{12}$CO.
This means that we can fit the observed $^{12}$CO intensity using either 
a lower disk mass with a smaller $\gamma$ or higher disk mass with bigger $\gamma$.
Because the inner disk region is optically thick, 
and the key factor to derive a good fit is to adjust the 
density profile in the outer disk region to obtain consistent intensity there.

4, The morphology of the $^{12}$CO channel maps are important constraints
of the disk mass. In our parameter space, although there are some models with 
higher or lower disk masses that can result in a reasonable reduced $\chi^2$,
their channel maps show a double cone structure that is 
either too strong or too weak compared to the ALMA observation.
The best-fit model with a disk mass of $\sim$ 0.1 $M_{\odot}$ can 
best reproduce the observed channel maps.

\acknowledgments
This paper makes use of the following ALMA data: ADS/JAO.ALMA\#2012.1.00870.S. ALMA is a partnership of ESO (representing its member states), NSF (USA) and NINS (Japan), together with NRC (Canada), MOST and ASIAA (Taiwan), and KASI (Republic of Korea), in cooperation with the Republic of Chile. The Joint ALMA Observatory is operated by ESO, AUI/NRAO and NAOJ.
S.J. and J.J. acknowledge support from the National Natural Science Foundation of China (grant Nos. 11503092, 11773081, 11661161013, 11573073, 11633009), the CAS Interdisciplinary Innovation Team, 
the Strategic Priority Research Program on Space Science, the Chinese Academy of Sciences, Grant No. XDA15020302 
and the Foundation of Minor Planets of Purple Mountain Observatory. 
A.I. acknowledges support from the
National Science Foundation through grant No. AST-1715719
and  the  National  Aeronautics  and  Space  Administration
support from the National Aeronautics and Space Administration through grant No. 80HQTR18T0061 and the Center for Space and Earth Science at LANL.
We thank the referee for comments that helped to improve the manuscript.

\software{RADMC-3D v0.41\citep{dullemond2012radmc}, CASA ALMA pipeline 5.4.0}

\begin{table} 
\centering
\caption{Four-dimensional parameter grids}
\begin{tabular}{lc}
\hline
\hline
Parameter & Grids \\
\hline
$M_{disk}$ ($M_{\odot}$) & [1.0e-4, 3.3e-4, 1.0e-3, 3.3e-3, 1.0e-2, 3.3e-2, 1.0e-1, 3.3e-1] \\
$\gamma$  & [0.5, 1.0, 2.0, 3.0, 4.0, 5.0, 6.0, 7.0] \\
$RC_{\rm arctan}$ &    [1, 5, 15,  25, 35, 45, 55, 65] \\
$\gamma_{\rm arctan}$ & [2, 4, 6, 8, 10, 12, 14, 16] \\
\hline
\vspace{0.05cm}
\end{tabular}
\label{para}
\end{table}

\end{document}